\documentclass[a4paper]{article}

\usepackage{INTERSPEECH2020}
\usepackage{multirow}
\usepackage{diagbox}
\usepackage{slashbox}
\usepackage{verbatim}
\usepackage{hyperref}
\usepackage{graphicx}
\usepackage[table,xcdraw]{xcolor}
\title{Towards Robust Neural Vocoding for Speech Generation: A Survey}
\name{Po-chun Hsu\textsuperscript{*}$^{1 2}$\thanks{*Equal contribution. } \qquad Chun-hsuan Wang\textsuperscript{*}$^{1 2}$\thanks{This work was supported by NVIDIA, TWCC, and Taiwan AI Labs.} \qquad Andy T. Liu$^{1 2}$ \qquad Hung-yi Lee$^{1 2}$}
\address{$^1$College of Electrical Engineering and Computer Science, National Taiwan University\\
        $^2$Graduate Institute of Communication Engineering, National Taiwan University}
\email{\{f07942095, r07942076, r07942089, hungyilee\}@ntu.edu.tw}

\begin{document}

\maketitle
\begin{abstract}
Recently, neural vocoders have been widely used in speech synthesis tasks, including text-to-speech and voice conversion. 
However, when encountering data distribution mismatch between training and inference,
neural vocoders trained on real data often degrade in voice quality for unseen scenarios.
In this paper, we train four common neural vocoders, including WaveNet, WaveRNN, FFTNet, Parallel WaveGAN alternately on five different datasets.
To study the robustness of neural vocoders, we evaluate the models using acoustic features from seen/unseen speakers, seen/unseen languages, a text-to-speech model, and a voice conversion model.
We found out that the speaker variety is much more important for achieving a universal vocoder than the language.
Through our experiments, we show that WaveNet and WaveRNN are more suitable for text-to-speech models, while Parallel WaveGAN is more suitable for voice conversion applications.
Great amount of subjective MOS results in naturalness for all vocoders are presented for future studies.
\end{abstract}

\section{Introduction}
\label{sec:intro}
Most speech generation models, such as text-to-speech \cite{ping2017deep, wang2017tacotron, shen2018natural, ren2019fastspeech} and voice conversion \cite{chou2018multi, pmlr-v97-qian19c, kameoka2018stargan, serra2019blow}, do not generate waveform directly.
Instead, the models output acoustic features such as Mel-spectrograms or F0 frequencies. Traditionally, waveform can be vocoded from these acoustic or linguistic features using heuristic methods \cite{griffin1984signal} or handcrafted vocoders \cite{morise2016world, kawahara2006straight, banno2007implementation}.
However, due to the assumptions under the heuristic methods, the quality of the generated speech is largely limited and undermined.

Since Tacotron 2 \cite{shen2018natural} first applied WaveNet \cite{oord2016wavenet} as a vocoder to generate waveform from Mel-spectrograms, neural vocoders have gradually become the most common vocoding method for speech synthesis.
Nowadays, neural vocoders have replaced traditional heuristic methods and dramatically enhance the quality of generated speech.
WaveNet generates waveform in high quality but costs long inference time due to the autoregressive architecture. 
To solve this problem, fast inference architecture, such as FFTNet \cite{jin2018fftnet}, WaveRNN \cite{kalchbrenner2018efficient}, LPCNet\cite{valin2019lpcnet}, and WaveGlow \cite{prenger2019waveglow},  have been proposed.

Neural vocoders can successfully model the data distribution of human voice with acoustic features \cite{oord2016wavenet, jin2018fftnet, kalchbrenner2018efficient, valin2019lpcnet, prenger2019waveglow}, however the generated speech quality is still restricted by the consistency of training and testing domain due to deep learning's data-driven property.
Recently, \cite{lorenzo2018towards} reported that a WaveRNN-based neural vocoder trained on multi-speaker multilingual data can generate natural speech despite conditions from an unseen domain.
However, there is still a lot to be studied about the robustness between different vocoders and their applications on various speech generation tasks.

In this paper, we survey a variety of neural vocoders trained on datasets across different domains applied to several scenarios. The contributions of this work are:
\begin{itemize}
\vspace{-3pt}
\item We construct 5 datasets, including single-speaker/mulit-speaker and monolingual/multilingual dataset, then alternately train 4 neural vocoder on 5 datasets to find how they perform when speakers and language are out of domain. 
\vspace{-3pt}
\item  The performances of neural vocoders are investigated by testing on human speech, voice conversion, and text-to-speech.
\vspace{-3pt}
\item  We analyze the robustness of neural vocoders in different scenarios based on mean opinion score (MOS) survey. 
\vspace{-3pt}
\end{itemize}{}

In Section \ref{sec:System}, we first introduce all the vocoder architectures used in this paper.
In Section \ref{sec:Datasets}, we introduce the datasets and specify the evaluation metrics.
We present three conducted experiments in Section \ref{sec:Robustness}, \ref{sec:gender}, \ref{sec:VC-TTS}.
In Section \ref{sec:Robustness}, we evaluate the trained vocoders on human speech.
In Section \ref{sec:gender}, we analyze the influence of the speaker's gender on vocoders.
In Section \ref{sec:VC-TTS}, we trained vocoders to speech synthesis tasks.
We then conclude our results in Section \ref{sec:Conclusion}.

\section{Neural Vocoders}\label{sec:System}
%
\subsection{WaveNet}
\label{ssec:WaveNet}
WaveNet \cite{oord2016wavenet} is an autoregressive model that directly generates audio samples. The network architecture is composed of layers of dilated causal convolution with gated activation units \cite{van2016conditional} for non-linearity. 
Our WaveNet model is modified from the public implementation \footnote{\url{https://github.com/r9y9/wavenet_vocoder}}, with 30 layers, 3 dilation cycles, 128 residual channels, 256 gate channels, and 128 skip channels. The input and output are 8-bit one-hot vectors quantized using µ-law companding transformation \cite{recommendation1988pulse}. We trained the model with a batch size of 6 on a single NVIDIA 1080Ti for 500k iterations, which takes 4 days to converge.
\subsection{WaveRNN}
\label{ssec:WaveRNN}
The output of original WaveRNN\cite{kalchbrenner2018efficient} is 16-bit quantized integer with two softmax predictions.  
To compare the quality between different vocoder models, our version of WaveRNN outputs 8-bit quantized integer with only a softmax prediction.
It can be seen as \cite{lorenzo2018towards} with only modification of internal layers.
The WaveRNN model we used is based on the public implementation \footnote{\url{https://github.com/fatchord/WaveRNN}}.  The conditioning module consists of upsampling layers with residual connections.  The network is trained with a batch size of 32 on a single NVIDIA V100 for 500k iterations and converged in 2 days.
\subsection{FFTNet}
\label{ssec:FFTNet}
The input of the original FFTNet\cite{jin2018fftnet} is Mel Cepstral Coefficients (MCC) and fundamental frequencies (F0). 
In this paper, to align the comparison between other vocoder models, we change the input into Mel-spectrogram. 
Inference techniques listed on the origin paper are added. 
The FFTNet model we used is a modification of the public implementation \footnote{\url{https://github.com/yoyololicon/pytorch_FFTNet}}.  
The network is trained with a batch size of 32 on a single NVIDIA V100 for 500k iterations and converged in 3 days.

\subsection{Parallel WaveGAN}
Parallel WaveGAN~\cite{yamamoto2020parallel} is a non-autoregressive neural vocoder trained to minimize the multi-resolution STFT loss and the waveform-domain adversarial loss.  The model synthesis speech in parallel with good quality.
We trained the parallel WaveGAN modified from the public implementation\footnote{\url{https://github.com/kan-bayashi/ParallelWaveGAN}} on an  NVIDIA V100, and it converged in 3 days.

\section{Datasets and Evaluation Metrics}
\label{sec:Datasets}
\subsection{Datasets for experiments}
\label{ssec:dataset_experiments}
The following datasets are used in our training: CMU US BDL Arctic Dataset (cmu\_ma), CMU US SLT Arctic Dataset (cmu\_fe)~\cite{john2004cmu}, Internal Mandarin Dataset (man\_fe), LibriTTS (libri)~\cite{zen2019libritts}, and Bible (bible).

CMU US BDL/SLT Arctic is a single English male/female speaker dataset.
Internal Mandarin is a single Mandarin female speaker  dataset.
We choose LibriTTS train\_claen as one of our dataset. 
Bible is collected from a Bible reading website\footnote{\url{http://www.bible.is}}.
The labels inside the bracket are the abbreviation of the datasets in Table \ref{tab:train_set}.
The sampling rate of all datasets are greater or equal to 22050 Hz, and we resample all dataset to 22050 Hz for experiments.

We compose the different listed datasets to our training set in Table \ref{tab:train_set}.
The number of speakers of Lrg is not a certain number since in bible there might be several speakers in one utterance.
The detail information of the testing data for Section~\ref{sec:Robustness} and~\ref{sec:gender} is listed in Table \ref{tab:test_set}.  
In the following sections, dataset starting with a capital letter informs training set, and dataset with Italic type informs testing set.

LJ Speech (lj) \cite{ljspeech17} is a commonly used dataset for training text-to-speech model and neural vocoders. It includes 13100 clean utterances recorded from a female English speaker. It is used to train our text-to-speech model and the baseline vocoder for the text-to-speech experiment in Section \ref{sec:VC-TTS}.

VCTK (vctk) \cite{vctk17} is a multi-speaker English dataset.  It is used to train our voice conversion model and the baseline vocoder for the voice conversion experiment in Section \ref{sec:VC-TTS}.

For all following experiments, We used the 80-band Mel-spectrogram as the auxiliary condition to synthesize audio. The FFT size, hop size and window size for STFT are 2048, 200, and 800, respectively.

\begin{table}[t]
\caption{Overview of the training datasets. }
\label{tab:train_set}
\centering
\vspace{-8pt}
\begin{tabular}{|c|l|c|c|l|}
\hline
\textbf{label}      &  \begin{tabular}[c]{@{}l@{}}\textbf{consist} \\ \textbf{datasets}\end{tabular}                                                                 & \textbf{speakers} & \textbf{utterances} & \begin{tabular}[c]{@{}l@{}}\textbf{consist} \\ \textbf{language}\end{tabular}                                                                                         \\\hline
En\_M   & cmu\_ma                                                                       & 1             & 1091             & English                                                                                          \\\hline
En\_F  & cmu\_fe                                                                     & 1             & 1092             & English                                                                                          \\\hline
Ma\_F  & man\_fe                                                                          & 1             & 8904             & Mandarin                                                                                         \\\hline
En\_L  & \begin{tabular}[c]{@{}l@{}}cmu\_ma\\ cmu\_fe\\ libri\end{tabular}         & 560           & 35419            & English                                                                                          \\\hline
Lrg & \begin{tabular}[c]{@{}l@{}}cmu\_ma\\ cmu\_fe\\ libri\\ bible\end{tabular} & \textgreater 600          & 38139            & \begin{tabular}[c]{@{}l@{}}English\\ French\\ Japanese\\ Korean\\ Spanish\\ Thai\end{tabular} \\ \hline
\end{tabular}
\vspace{-5pt}
\end{table}

\begin{table}[t]
\caption{Overview of testing data for Section \ref{sec:Robustness}, \ref{sec:gender}.}
\label{tab:test_set}
\centering
\vspace{-8pt}
\resizebox{80mm}{!}{
\begin{tabular}{|c|c|c|c|c|c|c|}
\hline
\multicolumn{2}{|c|}{\multirow{2}{*}{\begin{tabular}[c]{@{}c@{}}test set\\ label\end{tabular}}} & \multicolumn{2}{c|}{\# Speakers} & \multirow{2}{*}{\begin{tabular}[c]{@{}c@{}}Utter.\\ Num\end{tabular}} & \multirow{2}{*}{\begin{tabular}[c]{@{}c@{}}Speaker\\ same as\end{tabular}} & \multirow{2}{*}{MOS} \\ \cline{3-4}
\multicolumn{2}{|c|}{}                                                                          & F               & M              &                                                                       &                                                                            &                      \\ \hline
\multicolumn{2}{|c|}{ \emph{en\_m}}                                                                     & 0               & 1              & 10                                                                    & En\_M                                                                      & 4.79$\pm$0.10            \\ \hline
\multicolumn{2}{|c|}{\emph{en\_f}}                                                                     & 1               & 0              & 10                                                                    & En\_F                                                                      & 4.67$\pm$0.14           \\ \hline
\multicolumn{2}{|c|}{\emph{ma\_f}}                                                                     & 1               & 0              & 10                                                                    & Ma\_F                                                                      & 4.55$\pm$0.15           \\ \hline
\multirow{2}{*}{\emph{en\_l}}                                    & \emph{m}                                   & 0               & 10             & 10                                                                    & \multirow{4}{*}{\begin{tabular}[c]{@{}c@{}}No\\ overlap\end{tabular}}      & 4.64$\pm$0.13            \\ \cline{2-5} \cline{7-7} 
                                                          & \emph{f }                                  & 10              & 0              & 10                                                                    &                                                                            &   4.43$\pm$0.14         \\ \cline{1-5} \cline{7-7} 
\multirow{2}{*}{\emph{ma\_l}}                                    & \emph{m}                                   & 0               & 3              & 10                                                                    &                                                                            & 4.32$\pm$0.16            \\ \cline{2-5} \cline{7-7} 
                                                          & \emph{f }                                  & 3               & 0              & 10                                                                    &                                                                            & 4.54$\pm$0.15            \\ \hline
\end{tabular}
}
\vspace{-15pt}
\end{table}

\subsection{Evaluation metrics}
We conduct Mean Opinion Score (MOS) tests\footnote{Audio samples are publicly available at \url{https://bogihsu.github.io/Robust-Neural-Vocoding/}} to rate the quality of the generated speech. Each utterance was scored based on its naturalness on a 1-to-5 scale. A higher score signifies a more natural utterance.
All of the MOS results are reported with 95\% confidence intervals.  Each score is conducted of 10 utterances; each utterance was rated by at least 10 raters. 
More than 350 subjects were surveyed in the experiments for Section~\ref{sec:Robustness} and~\ref{sec:gender}, and more than 120 for Section~\ref{sec:VC-TTS}.
Evaluations for the ground truth in testing data were conducted together with the experiments in Section~\ref{sec:Robustness} and~\ref{sec:gender}.
The results are listed in Table~\ref{tab:test_set}.

\section{Robustness to Human Speech}\label{sec:Robustness}
In this section, we consider synthesizing speech conditioned on Mel-spectrograms extracted from the ground truth data.
\subsection{Experimental setup}
\label{ssec:ES1}
To observe how the vocoder models perform when facing inconsistent train/test scenarios,
Vocoder models are tested on seen/unseen speakers and seen/unseen languages settings after training, where training set composed in Table \ref{tab:train_set}.
However, there do not exist any dataset with a speaker that speak more than one language, the testing scenario with seen speaker and unseen language can't be tested.

Therefore, for each trained vocoder model, it is tested in 3 situations, including seen speakers seen languages/unseen speakers seen languages/unseen speakers unseen language.
To sum up, for 4 vocoder models, WaveNet, WaveRNN, FFTNet, and Parallel WaveGAN, are tested in 15 scenarios (5 train sets $\times$ 3 situations)  listed in Table \ref{tab:scenario_lan_test}, where SS/US/UU correspond to Seen speakers Seen languages/Unseen speakers Seen languages/Unseen speakers Unseen languages.

For the scenario of seen speakers and seen languages (SS), vocoders are trained and evaluated on training and testing data from the same datasets.
Since the training data En\_L and Lrg both contain En\_F and En\_M, therefore we choose testing data from en\_f and en\_m.  
The detail statistics of test set are listed in Table \ref{tab:test_set}.  
For those having half label, we only choose half of the utterances to match number the testing utterances for all scenarios.

For the scenario of unseen speakers and seen languages (US), vocoders are tested with a multi-speaker dataset from the same language with no speakers overlapping between training and testing. 
For the scenario of unseen speakers and unseen languages (UU), vocoders are tested with a multi-speaker dataset on a unseen language. There are no speakers overlapping between training and testing as well.


%

\begin{table}[t]
\centering
\caption{Scenario of testing the influence for  seen/unseen speakers and seen/unseen language}
\label{tab:scenario_lan_test}
\vspace{-8pt}
\begin{tabular}{|c|c|c|c|c|c|}
\hline
\multirow{2}{*}{\begin{tabular}[c]{@{}l@{}}Train Set\\ Label\end{tabular}} & \multicolumn{2}{c|}{Train Set Char.}             & \multicolumn{3}{c|}{Correspond Test Set}                                                                                                                                                                   \\ \cline{2-6} 
                                                                           & Speaker                 & lingual                & SS         & US & UU \\ \hline
Ma\_F                                                                      & \multirow{3}{*}{single} & \multirow{4}{*}{mono-} & \emph{ma\_f }                                                                 & \emph{ma\_l}/2                                                           & \emph{en\_l}/2                                                           \\ \cline{1-1} \cline{4-6} 
En\_M                                                                      &                         &                        & \emph{en\_m}                                                                  & \multirow{4}{*}{\emph{en\_l}/2}                                          & \multirow{4}{*}{\emph{ma\_l}/2}                                          \\ \cline{1-1} \cline{4-4}
En\_F                                                                      &                         &                        & \emph{en\_f}                                                                  &                                                                 &                                                                 \\ \cline{1-2} \cline{4-4}
En\_L                                                                      & \multirow{2}{*}{multi}  &                        & \multirow{2}{*}{\begin{tabular}[c]{@{}l@{}}\emph{en\_m}/2\\ \emph{en\_f}/2\end{tabular}} &                                                                 &                                                                 \\ \cline{1-1} \cline{3-3}
Lrg                                                                        &                         & mulit-                 &                                                                        &                                                                 &                                                                 \\ \hline
\end{tabular}
\vspace{-5pt}
\end{table}

\begin{table}[!t]
\caption{MOS for Speaker and language in/out domain experiments}
\label{tab:lan_test}
\centering
\vspace{-8pt}
\resizebox{80mm}{!}{
\begin{tabular}{cccccc}
\toprule
   & \multicolumn{5}{c}{Vocoder Training Set} \\
   & En\_F  & En\_M  & Ma\_F  & En\_L  & Lrg  \\ \bottomrule
\multicolumn{6}{c}{Seen Speakers and Seen Language}          \\ \hline
WN & \textbf{4.78$\pm$0.10} & \textbf{4.71$\pm$0.11} & 4.63$\pm$0.12          & \textbf{4.72$\pm$0.10} & \textbf{4.70$\pm$0.13} \\
\rowcolor[HTML]{EFEFEF}
WR & 4.48$\pm$0.13          & 4.61$\pm$0.13          & \textbf{4.66$\pm$0.11} & 4.64$\pm$0.11          & 4.61$\pm$0.13          \\
FF & 3.87$\pm$0.17          & 4.29$\pm$0.15          & 4.45$\pm$0.10          & 3.28$\pm$0.19          & 3.58$\pm$0.17          \\
\rowcolor[HTML]{EFEFEF}
PW & 4.59$\pm$0.12          & 4.29$\pm$0.17          & 4.41$\pm$0.12          & 4.29$\pm$0.15          & 4.11$\pm$0.16          \\ \bottomrule
\multicolumn{6}{c}{Unseen Speakers and Seen Language}                        \\ \hline
WN & 2.27$\pm$0.14          & 2.86$\pm$0.17          & 3.27$\pm$0.16          & \textbf{4.25$\pm$0.17} & \textbf{4.35$\pm$0.15} \\
\rowcolor[HTML]{EFEFEF}
WR & \textbf{2.60$\pm$0.14} & \textbf{2.89$\pm$0.15} & \textbf{3.54$\pm$0.14} & 3.98$\pm$0.15          & 3.92$\pm$0.16          \\
FF & 1.76$\pm$0.15          & 2.21$\pm$0.14          & 2.94$\pm$0.13          & 2.99$\pm$0.18          & 3.13$\pm$0.21          \\
\rowcolor[HTML]{EFEFEF}
PW & 2.35$\pm$0.15          & 2.85$\pm$0.16          & 2.88$\pm$0.14          & 3.80$\pm$0.21          & 3.85$\pm$0.17          \\ \bottomrule
\multicolumn{6}{c}{Unseen Speakers and Unseen Language}                        \\ \hline
WN & 1.90$\pm$0.12          & 2.53$\pm$0.12          & \textbf{3.85$\pm$0.15} & \textbf{4.33$\pm$0.15} & \textbf{4.33$\pm$0.17} \\
\rowcolor[HTML]{EFEFEF}
WR & \textbf{2.53$\pm$0.13} & \textbf{2.62$\pm$0.12} & 3.30$\pm$0.15          & 4.30$\pm$0.16          & 4.16$\pm$0.17          \\
FF & 1.56$\pm$0.09          & 1.75$\pm$0.12          & 2.64$\pm$0.16          & 2.67$\pm$0.17          & 3.37$\pm$0.17       \\
\rowcolor[HTML]{EFEFEF}
PW & 2.17$\pm$0.11          & 2.54$\pm$0.12          & 2.49$\pm$0.13          & 3.79$\pm$0.20          & 3.97$\pm$0.19  \\  \bottomrule
\end{tabular}
}
\vspace{-15pt}
\end{table}

\subsection{Results}\label{ssec:Res1}
All training criteria are listed in Section \ref{sec:System}.
The results are listed in Table \ref{tab:lan_test}, where WN, WR, FF, PW represent WaveNet, WaveRNN, FFTNet, Parallel WaveGAN, respectively.

\subsubsection{Seen Speakers and Seen language}
From the 1st block in Table \ref{tab:lan_test}, for testing data contains  seen speakers and seen language from training data, WaveNet perform the best of all, some even better than the ground truth test data.  
We suppose that the ground truth data may have a little microphone background noise, where WaveNet model can eliminate a little.
All models perform quite well in the same domain of seen speaker and seen language.

\subsubsection{Unseen speakers and seen language}
In the 2nd block of Table \ref{tab:lan_test}, there is a gap between seen speakers and unseen speakers for all vocoder models, especially in single speaker dataset.
Trained in a single speaker dataset, the results of all models are staticky.
However, with huge amount of data, the performance degradation will be relieved.

For all vocoder models,the WaveNet model has stronger robustness for out-of-domain speakers when it is in a multi-speaker dataset, while the WaveRNN model has stronger robustness when it is trained in a single-speaker dataset.
Furthermore, when we compare models trained dataset En\_L and Lrg, we found that the larger the training data is the better it is preformed for FFTNet and Parallel WaveGAN, while the WaveNet and WaveRNN perform very similar.

\subsubsection{Unseen speakers and unseen language}
For unseen speakers and unseen language, vocoder models' performances are comparable to the case of unseen speakers and seen language.
Hence, we conclude that the robustness for a vocoder is caused by the speaker variety. 
With enough training data, vocoder models can perform similar regardless the language is in/out of training domain.

\subsubsection{Discussion}
Trained in a multi-speaker dataset, vocoder models can perform very similar regardless the in/out of domain speakers and language.
Language out of domain doses not influence the model performance.
On the contrary, speakers out of domain influence very much.
With large variety of the training data can help set up a universal vocoder.

\section{The Influence of Genders}\label{sec:gender}
In Section \ref{sec:Robustness}, we survey how unseen speakers influence the neural vocoder models. 
However, for vocoders trained on single speaker dataset, we cannot figure the degradation of test performance unseen speakers is caused by unseen speakers or unseen gender.

To investigate more, we conduct the following experiment to explore how vocoder's behaviour is influenced by speaker gender.

\subsection{Experimental setup}
\label{ssec:ES2}
In this section, to discuss model sensitivity on unseen gender, neural vocoders trained on single speaker datasets (e.g. En\_M, En\_F, Ma\_F) will be considered.  The model will be tested on unseen speakers to find out the influence of genders.  The scenario for the training and testing are listed in Table \ref{tab:gender_test_scenario}, where  SS/US/SU/UU  correspond to Seen gender Seen languages/Unseen gender Seen languages/Seen gender Unseen languages/Unseen gender Unseen languages.

\begin{table}[!h]
\vspace{-5pt}
\caption{Scenario of testing the influence for seen/unseen gender and seen/unseen language}
\label{tab:gender_test_scenario}
\centering
\vspace{-8pt}
\begin{tabular}{|c|c|c|c|c|}
\hline
\multirow{2}{*}{\begin{tabular}[c]{@{}c@{}}Train Set\\ Label\end{tabular}} & \multicolumn{4}{c|}{Correpsond Test Set Label} \\ \cline{2-5} 
                                                                           & SS         & US        & SU        & UU        \\ \hline
En\_M                                                                      & \emph{en\_l\_m}      & \emph{en\_l\_f}     & \emph{ma\_l\_m}     & \emph{ma\_l\_f}     \\ \hline
En\_F                                                                      & \emph{en\_l\_f}      & \emph{en\_l\_m}     & \emph{ma\_l\_f}     & \emph{ma\_l\_m}     \\ \hline
Ma\_F                                                                      & \emph{ma\_l\_m}      & \emph{ma\_l\_m}     & \emph{en\_l\_f}     & \emph{en\_l\_m}     \\ \hline
\end{tabular}
\vspace{-15pt}
\end{table}

\subsection{Results}
\label{ssec:Res2}

The results are listed in Table \ref{tab:gend_test}.  The generalization capability  trained in a single speaker is worse than those trained with multiple speakers for all models.
Trained with a female speaker, models tend to perform better in average for all models.
However, when tested in male dataset, those trained in a female speaker dataset still cannot beat trained in a male speaker dataset.
Hence, both female and male speakers are essential in training to have a nice and descent result. 

\begin{table}[!t]
\caption{MOS for Gender and language in/out domain experiments}
\label{tab:gend_test}
\centering
\vspace{-8pt}
\resizebox{75mm}{!}{
\begin{tabular}{cccccc}
\toprule
\multirow{2}{*}{Model}   & \multicolumn{3}{c}{Vocoder Training Set} \\
   & En\_M  & En\_F  & Ma\_F   \\ \bottomrule
\multicolumn{4}{c}{Seen Gender and Seen Language}          \\ \hline
WaveNet          & 2.41$\pm$0.23 & 3.47$\pm$0.24 & 3.57$\pm$0.20 \\
\rowcolor[HTML]{EFEFEF}
WaveRNN          & 2.85$\pm$0.21 & 3.49$\pm$0.21 & 4.08$\pm$0.20 \\
FFTNet           & 2.01$\pm$0.24 & 2.45$\pm$0.21 & 3.56$\pm$0.14 \\
\rowcolor[HTML]{EFEFEF}
Parallel WaveGAN & 2.68$\pm$0.22 & 3.47$\pm$0.20 & 3.34$\pm$0.17 \\  \bottomrule
\multicolumn{4}{c}{Unseen Gender and Seen Language}          \\ \hline
WaveNet          & 2.13$\pm$0.16 & 2.25$\pm$0.16 & 2.98$\pm$0.21 \\
\rowcolor[HTML]{EFEFEF}
WaveRNN          & 2.36$\pm$0.20 & 2.29$\pm$0.15 & 3.01$\pm$0.20 \\
FFTNet           & 1.52$\pm$0.15 & 1.97$\pm$0.20 & 2.34$\pm$0.15 \\
\rowcolor[HTML]{EFEFEF}
Parallel WaveGAN & 2.03$\pm$0.17 & 2.23$\pm$0.18 & 2.41$\pm$0.17 \\\bottomrule
\multicolumn{4}{c}{Seen Gender and Unseen Language}                        \\ \hline
WaveNet          & 1.92$\pm$0.16 & 3.05$\pm$0.23 & 4.10$\pm$0.22 \\
\rowcolor[HTML]{EFEFEF}
WaveRNN          & 2.78$\pm$0.18 & 3.12$\pm$0.21 & 3.77$\pm$0.18 \\
FFTNet           & 1.74$\pm$0.17 & 2.00$\pm$0.17 & 3.40$\pm$0.17 \\
\rowcolor[HTML]{EFEFEF}
Parallel WaveGAN & 2.29$\pm$0.19 & 2.92$\pm$0.22 & 2.92$\pm$0.21 \\ \bottomrule
\multicolumn{4}{c}{Unseen Gender and Unseen Language}                        \\ \hline
WaveNet          & 1.88$\pm$0.16 & 2.01$\pm$0.16 & 3.59$\pm$0.20 \\
\rowcolor[HTML]{EFEFEF}
WaveRNN          & 2.29$\pm$0.17 & 2.12$\pm$0.19 & 2.84$\pm$0.21 \\
FFTNet           & 1.38$\pm$0.11 & 1.51$\pm$0.11 & 1.91$\pm$0.16 \\
\rowcolor[HTML]{EFEFEF}
Parallel WaveGAN & 2.06$\pm$0.16 & 2.17$\pm$0.15 & 2.05$\pm$0.17  \\  \bottomrule
\end{tabular}
}
\vspace{-15pt}
\end{table}

\section{Robustness to Speech Synthesis Task}
\label{sec:VC-TTS}
The neural vocoder was originally proposed as a vocoder for the text-to-speech model \cite{shen2018natural}. In this section, we test the performances of vocoders by applying them to speech synthesis tasks.  Both implementation of the text-to-speech\footnote{\url{https://github.com/NVIDIA/tacotron2}} and voice conversion\footnote{\url{https://github.com/BogiHsu/Voice-Conversion}} are publicly available.
\subsection{Experimental setup}
\label{ssec:ES3}
Neural vocoders are more frequently used to generate audio from the output of upstream speech tasks, such as text-to-speech synthesis model or voice conversion model.  Hence, experiments are examined to find out which model can perform better.

\subsubsection{Text-to-speech synthesis}
Tacotron 2 \cite{shen2018natural} is examined for text-to-speech synthesis, and was trained on LJ Speech \cite{ljspeech17}.
Vocoders trained on LJ Speech is the topline model.
The Mel-spectrograms generated by the Tacotron 2 are fed to the vocoders trained with different datasets listed in Table \ref{tab:train_set}. 
We also trained a vocoder on ground-truth aligned predictions~\cite{shen2018natural} from the 
Tacotron 2, which is noted as Cond in Table~\ref{tab:tts_test}.

\subsubsection{Voice conversion}
We examined the voice conversion model in \cite{chou2018multi}.
The voice conversion model was trained on VCTK, hence the vocoders trained on the same dataset are the topline model.
The output of the voice conversion model is linear scale and fed to a Mel-filter to get Mel-spectrogram.  
For comparison, We also tested a heuristic method, Griffin-Lim algorithm (GL)~\cite{griffin1984signal}, which reconstructs signals directly from the linear spectrograms.  

\subsection{Results}
\label{ssec:Res3}
\subsubsection{Text-to-speech synthesis}
We compare the text-to-speech result fed to vocoders and the same utterances from LJ Speech in Table \ref{tab:tts_test}, where WN, WR, FF, PW, GT represent WaveNet, WaveRNN, FFTNet, Parallel WaveGAN, ground truth, respectively.

The upper bound of the text-to-speech model is the result of the Condition model.  However, WaveRNN and Parallel WaveGAN do not perform the best in all training set.  We suppose that the clear data are more important for it training stability.
Both WaveNet, WaveRNN model trained in the same source of text-to-speech system, LJ Speech, perform very clearly and natural in the experiment result.

\subsubsection{Voice conversion}
We compare the voice conversion result fed to vocoders and Griffin-Lim algorithm in Table \ref{tab:vc_test}, where WN, WR, FF, PW, GL represent WaveNet, WaveRNN, FFTNet, Parallel WaveGAN, Griffin-Lim algorithm, respectively.

%
The result indicates that neural vocoders outperform the Griffin-Lim algorithm regardless of the training data.
Particularly, Parallel WaveGAN performs best in naturalness over the other competitors.
Hence, for application usage, Parallel WaveGAN is recommended to be used as a vocoder and trained on the same dataset used for training the voice conversion model.
Great amount of data are also recommended for Parallel WaveGAN vocoder to construct a universal vocoder for voice conversion experiments.

\begin{table}[t]
\caption{MOS for text-to-speech (TTS) synthesis experiment}
\label{tab:tts_test}
\centering
\vspace{-8pt}
\resizebox{80mm}{!}{
\begin{tabular}{cccccc}
\toprule
   & \multicolumn{5}{c}{Vocoder Training Set} \\
   & LJ  & En\_F  &  En\_L  & Lrg  & Cond  \\ \bottomrule
WN & 4.10$\pm$0.19 & 2.59$\pm$0.24 & 3.54$\pm$0.20 & 3.66$\pm$0.21 & \textbf{4.21$\pm$0.16} \\
\rowcolor[HTML]{EFEFEF}
WR & \textbf{4.16$\pm$0.18} & 3.05$\pm$0.24 & 3.32$\pm$0.20 & 3.73$\pm$0.19 & 3.79$\pm$0.19 \\
FF & 2.75$\pm$0.27 & 2.16$\pm$0.29 & 2.50$\pm$0.27 & 2.28$\pm$0.28 & 2.86$\pm$0.30 \\
\rowcolor[HTML]{EFEFEF}
PW & 3.81$\pm$0.20 & 3.17$\pm$0.21 & 3.60$\pm$0.20 & 3.19$\pm$0.20 & 3.38$\pm$0.20 \\ \hline
GT                   & \multicolumn{5}{c}{4.54$\pm$0.16}   \\ \bottomrule   
\end{tabular}
}
\vspace{-8pt}
\end{table}

\begin{table}[t]
\caption{MOS for voice conversion experiments}
\label{tab:vc_test}
\centering
\vspace{-8pt}
\resizebox{80mm}{!}{
\begin{tabular}{cccccc}
\toprule
   & \multicolumn{5}{c}{Vocoder Training Set} \\
   & VCTK  & En\_M & En\_F  &  En\_L  & Lrg   \\ \bottomrule
WN & 3.15$\pm$0.21 & 3.25$\pm$0.23 & 2.86$\pm$0.25 & 2.85$\pm$0.19 & 2.81$\pm$0.21 \\
\rowcolor[HTML]{EFEFEF}
WR & 3.54$\pm$0.20 & 3.21$\pm$0.23 & 2.98$\pm$0.23 & 2.88$\pm$0.22 & 2.90$\pm$0.21 \\
FF & 2.71$\pm$0.22 & 2.19$\pm$0.21 & 2.30$\pm$0.23 & 2.28$\pm$0.23 & 2.51$\pm$0.21 \\
\rowcolor[HTML]{EFEFEF}
PW & \textbf{3.83$\pm$0.20} & 3.30$\pm$0.23 & 3.02$\pm$0.24 & \textbf{3.45$\pm$0.20} & 3.40$\pm$0.21 \\ \hline
GL & \multicolumn{5}{c}{2.72$\pm$0.21} \\ \bottomrule  
\end{tabular}
}
\vspace{-15pt}
\end{table}
\section{Conclusion}
\label{sec:Conclusion}
By tested on human speech, we conclude that
the speaker variety is more important than the language, when encountering unseen speaker and unseen language in testing.  
In total, the WaveNet model is more robust when encountering inconsistency between training data and testing data for most cases.  However, it has the slowest inference time for all.
The WaveRNN model performs well in the same domain on training and testing. It is also a great option for text-to-speech synthesis.
The FFTNet model is a usable vocoder in the in domain data, but not as a good choice for a universal vocoder.
The Parallel WaveGAN model output has lower quality than the WaveNet and WaveRNN in human speech, but perform the best in voice conversion experiments.

\bibliographystyle{IEEEtran}

\bibliography{mybib}


\end{document}